\documentclass[english]{revtex4-1}
\usepackage[T1]{fontenc}
\usepackage[latin9]{inputenc}
\usepackage{amsmath}
\usepackage{amssymb}

\makeatletter
\usepackage{babel}

\makeatother

\usepackage{babel}
\begin{document}

\title{Relativistic probability amplitudes II. The photon}

\author{Scott E. Hoffmann}

\address{School of Mathematics and Physics,~~\\
 The University of Queensland,~~\\
 Brisbane, QLD 4072~~\\
 Australia}
\email{scott.hoffmann@uqconnect.edu.au}

\begin{abstract}
We identify momentum/helicity probability amplitudes for the photon
and find their relativistic transformation properties. We also find
their behaviour under space inversion and time reversal. The discussion
begins with a review of the unitary, irreducible representations of
the Poincare group for massless particles. The little group, the set
of Lorentz transformations that leave the momentum unchanged, is distinctly
different for massless particles compared to the little group of rest
frame rotations for a massive particle. We give a physical interpretation
of the little group for a massless particle. The explicit forms of
the Wigner rotations for general rotations and boosts are given. In
normalized superpositions of the basis vectors, we identify the momentum/helicity
probability amplitudes, show their probability interpretation and
find their transformation properties. We see that position eigenvectors
for the photon are not possible, not because of their masslessness
but because of their limited helicity spectrum. Instead, to have a
measure of localization for the photon, we use the expectations of
the electromagnetic field strength operators in a coherent state with
a mean photon number of unity. In the construction of the polarization
vectors appearing in the field strengths, we see the connection between
gauge invariance and Lorentz covariance.
\end{abstract}
\maketitle

\section{Introduction}

In the first paper of this series \cite{Hoffmann2018d}, this author
found two types of probability amplitudes for massive particles: momentum/spin-component
probability amplitudes and position/spin-component probability amplitudes.
These amplitudes have the same probability interpretations as their
counterparts in the nonrelativistic theory. The transformations of
these amplitudes under spacetime translations, rotations, boosts,
space inversion and time reversal were given. These are unitary transformations
(antiunitary in the case of time reversal) in that they preserve the
modulus-squared of scalar products. The result is a relativistic theory
of free massive particles with spin. We noted in that paper that the
transformation properties should be called nonmanifest covariance,
distinguished from the manifest covariance of scalars, four-vectors
and tensors.

In this paper, we turn our attention to massless particles in general
and the photon in particular. Here, we will identify momentum/helicity
probability amplitudes with a probability interpretation and derive
their transformation properties. Again, the covariance will be nonmanifest.
For the particular case of the photon, we will find that no position
eigenvectors satisfying the requirements of Newton and Wigner \cite{Newton1949}
can be constructed. The observable fact that photons can be at least
partially localized will lead us to find another measure of that localization
in particular expectation values of the electromagnetic field strength
operators.

In Section II we review the little group for massless particles, the
group of Lorentz transformations that leave the momentum unchanged.
This little group is the direct product of the subgroup of rotations
about the momentum direction and another subgroup of Lorentz transformations.
We provide a physical interpretation of these latter elements in terms
of boosts that leave energy unchanged followed by rotations.

In Section III we show the general transformation properties of the
eigenvectors of momentum and helicity that carry the unitary, irreducible
representations. We provide explicit forms of the Wigner rotation
angles for general rotations and boosts. As we did for massive particles
in \cite{Hoffmann2018d}, we construct normalized coherent superpositions
of the basis vectors. From the complex coefficients we identify the
momentum/helicity probability amplitudes. Their transformation properties
are then derived.

In Section IV we review the reason why there can be no position eigenvectors
for the photon. To find at least a partial measure of localization,
we consider expectations of the electromagnetic field strength operators
(from quantum electrodynamics \cite{Itzykson1980}) in a specially
constructed coherent state. To find the form of the field strengths
and their transformation properties, it is necessary to construct
four-component polarization vectors $\epsilon^{\mu}(k,\lambda).$
It will be seen that these Lorentz transform covariantly only up to
a gauge transformation. However the tensor appearing in the field
strengths is gauge invariant, with the consequence that the field
strengths transform locally as an antisymmetric tensor. In this Section
we also discuss the work of Sipe \cite{Sipe1995} and Bia\l ynicki-Birula
\cite{Bialynicki-Birula1994}.

Throughout this paper we use Heaviside-Lorentz units, in which $\hbar=c=\epsilon_{0}=\mu_{0}=1$.
We use the active convention for Poincaré transformations, in which,
for example, the boost by velocity $\boldsymbol{\beta}_{0}$ of a
particle of mass $m_{0}$ at rest produces a particle with momentum
$p^{\mu}=m_{0}(\gamma_{0},\gamma_{0}\boldsymbol{\beta}_{0})^{\mu},$
with $\gamma_{0}=1/\sqrt{1-\boldsymbol{\beta}_{0}^{2}}.$

\section{The little group for massless particles}

The little group of a particle is defined as the subgroup of the Lorentz
transformations that leave the momentum unchanged. For a massive particle,
the little group is clearly the set of all rotations in the rest frame.
Then we know that the finite-dimensional, unitary, irreducible representations
can be labelled by any spin $s=0,\frac{1}{2},1,\frac{3}{2},\dots$
The half-integral spins carry double-valued representations of rest
frame rotations.

The state vectors for a particle at rest, $|\,(m_{0},\boldsymbol{0}),s,m\,\rangle,$
are taken as the reference state vectors. Different values of $m$
are connected using the spin raising and lowering operators, which
have well-defined matrix elements in the Condon-Shortley phase convention
\cite{Messiah1961}. Then state vectors for general momentum, $p^{\mu}=(\omega,\boldsymbol{p})^{\mu},$
are obtained by the boost
\begin{equation}
\Lambda[p]\equiv\Lambda(\frac{\boldsymbol{p}}{\omega}),\label{eq:1}
\end{equation}
where the latter are functions of the boost velocity. Then, for example,
the transformation of this state vector under general boosts, $\Lambda,$
is obtained from
\begin{equation}
\Lambda\Lambda[p]=\Lambda[\Lambda p]\{\Lambda^{-1}[\Lambda p]\Lambda\Lambda[p]\}.\label{eq:2}
\end{equation}
The transformation in braces is a little group transformation, a Wigner
rotation in the rest frame.

For a massless particle, one subgroup of the little group is the set
of all rotations about the momentum direction. Since no rest state
is available, we take as the reference state vector one with four-momentum
$k_{0}^{\mu}=(\kappa,0,0,\kappa)^{\mu},$ with a particular energy,
$\kappa,$ and momentum in the $+\hat{\boldsymbol{z}}$ direction.
The finite-dimensional, unitary irreducible representations are labelled
by the helicity, $\lambda$, for $\lambda=0,\frac{1}{2},1,\frac{3}{2},\dots.$
The helicity is a Lorentz invariant, with different helicities coupled
only by space inversion. The reference state vector rotates as
\begin{equation}
U(R_{z}(\gamma))\,|\,k_{0},\lambda\,\rangle=|\,k_{0},\lambda\,\rangle e^{-i\lambda\gamma}.\label{eq:3}
\end{equation}
(Helicity eigenvectors can also be defined for massive particles,
and they behave this way under rotations about the momentum direction.)
For the photon, there are two possible helicities, $\lambda=\pm1,$
which correspond to left and right circular polarization, respectively
\cite{Jackson1975}. Linear polarizations can be constructed as linear
superpositions of these.

For a massless particle but not a massive particle, there are boosts
that change the momentum direction but leave the energy unchanged.
If we follow this boost by a rotation to bring the momentum back to
the $z$ direction, we have an example of the other little group elements
for massless particles. (They do form a group.) We call this product
of an isoenergetic boost and a rotation an IBR. The full little group
is the direct product of this subgroup and the rotations about the
momentum direction.

We find that if the boost direction has spherical polar angles $(\theta,\varphi)$,
then the isoenergetic boost velocity must be
\begin{equation}
\boldsymbol{\beta}_{0}(\theta,\varphi)=-\frac{2\cos\theta}{1+\cos^{2}\theta}\hat{\boldsymbol{\beta}}_{0}(\theta,\varphi).\label{eq:4}
\end{equation}
Note that the speed as written is less than unity on $0<\theta<\pi$
and is negative on $0<\theta<\pi/2.$ The final polar angle of the
boosted momentum is
\begin{equation}
\psi_{0}(\theta)=2\theta-\pi,\label{eq:5}
\end{equation}
with $-\pi<\psi(\theta)<\pi.$

If we define
\begin{equation}
\boldsymbol{\alpha}\equiv-2\cot\theta\,(\cos\varphi\,\hat{\boldsymbol{x}}+\sin\varphi\,\hat{\boldsymbol{y}}),\label{eq:6}
\end{equation}
for an isoenergetic boost in the plane of $\hat{\boldsymbol{z}}$
and $\hat{\boldsymbol{u}}_{1}=\cos\varphi\,\hat{\boldsymbol{x}}+\sin\varphi\,\hat{\boldsymbol{y}}$,
we find that the IBR has the Lorentz transformation matrix
\begin{equation}
\mathcal{L}_{\phantom{\mu}\nu}^{\mu}(\boldsymbol{\alpha})=[R(-\psi_{0}(\theta)\hat{\boldsymbol{u}}_{2})\Lambda(\boldsymbol{\beta}_{0}(\theta,\varphi))]_{\phantom{\mu}\nu}^{\mu}=\begin{pmatrix}1+\frac{1}{2}\boldsymbol{\alpha}^{2} & \alpha_{x} & \alpha_{y} & -\frac{1}{2}\boldsymbol{\alpha}^{2}\\
\alpha_{x} & 1 & 0 & -\alpha_{x}\\
\alpha_{y} & 0 & 1 & -\alpha_{y}\\
\frac{1}{2}\boldsymbol{\alpha}^{2} & \alpha_{x} & \alpha_{y} & 1-\frac{1}{2}\boldsymbol{\alpha}^{2}
\end{pmatrix}_{\phantom{\mu}\nu}^{\mu},\label{eq:7}
\end{equation}
where $\hat{\boldsymbol{u}}_{2}=\hat{\boldsymbol{u}}_{1}\times\hat{\boldsymbol{z}}$
and $R(\boldsymbol{\Omega})$ represents a rotation by angle $\Omega$
about the axis $\hat{\boldsymbol{\Omega}}.$

From this form, we can derive the group multiplication laws
\begin{align}
\mathcal{L}(\boldsymbol{\alpha}_{1})\mathcal{L}(\boldsymbol{\alpha}_{2}) & =\mathcal{L}(\boldsymbol{\alpha}_{1}+\boldsymbol{\alpha}_{2}),\nonumber \\
R_{z}(\gamma)\mathcal{L}(\boldsymbol{\alpha})R_{z}^{-1}(\gamma) & =\mathcal{L}(R_{z}(\gamma)\boldsymbol{\alpha}).\label{eq:8}
\end{align}
So we see that the little group is isomorphic to the group of translations
and rotations in a plane, the Euclidean group in two dimensions. It
can be shown that the two commuting generators of the IBRs, with respect
to $\alpha_{x}$ and $\alpha_{y},$ are
\begin{eqnarray}
L_{x} & = & K_{x}-J_{y},\nonumber \\
L_{y} & = & K_{y}+J_{x},\label{eq:9}
\end{eqnarray}
in terms of the boost generators, $\boldsymbol{K},$ and the angular
momenta, $\boldsymbol{J}.$ The generator of the $z$ rotations with
respect to the rotation angle is $J_{z}.$

Since we believe a photon is completely characterized by its momentum
and its helicity, the IBRs must be represented by unity acting on
the basis vectors $|\,k_{0},\lambda\,\rangle$. The procedure for
constructing momentum/helicity eigenvectors of general momentum, $k^{\mu}=(\omega,\boldsymbol{k})^{\mu},$
with $k^{2}=0$ and $k^{0}=\omega>0,$ is to first boost the reference
state by
\begin{equation}
\boldsymbol{\beta}(\omega,\kappa)=\frac{\omega^{2}-\kappa^{2}}{\omega^{2}+\kappa^{2}}\hat{\boldsymbol{z}}\label{eq:10}
\end{equation}
to produce energy $\omega,$ with the transformation denoted $\Lambda_{z}(\omega,\kappa),$
then rotate into the direction $\hat{\boldsymbol{k}}=(\theta_{k},\varphi_{k})$
using the standard rotation
\begin{equation}
R_{0}[\hat{\boldsymbol{k}}]=R_{z}(\varphi_{k})R_{y}(\theta_{k})R_{z}(-\varphi_{k}).\label{eq:11}
\end{equation}
This is
\begin{equation}
|\,k,\lambda\,\rangle=U(R_{0}[\hat{\boldsymbol{k}}])U(\Lambda_{z}(\omega,\kappa))\,|\,k_{0},\lambda\,\rangle=U(L(k,k_{0}))\,|\,k_{0},\lambda\,\rangle.\label{eq:12}
\end{equation}
All transformations of the resulting $|\,k,\lambda\,\rangle$ then
follow using similar arguments to those of the massive case, by identifying
products of transformations that are little group elements.

\section{Momentum-helicity probability amplitudes}

We summarize the Poincaré and inversion transformations for the photon,
which have been derived elsewhere \cite{Halpern1968}. The use of
the invariant normalization,
\begin{equation}
\langle\,k_{1},\lambda_{1}\,|\,k_{2},\lambda_{2}\,\rangle=\delta_{\lambda_{1}\lambda_{2}}\omega_{1}\delta^{3}(\boldsymbol{k}_{1}-\boldsymbol{k}_{2}),\label{eq:13}
\end{equation}
makes the following results take their simplest forms. We note that
helicity is invariant under all such transformations except space
inversion.
\begin{eqnarray}
\mathrm{Spacetime\ translations:}\quad U(T(a)\,|\,k,\lambda\,\rangle & = & |\,k,\lambda\,\rangle\,e^{+ik\cdot a},\nonumber \\
\mathrm{Rotations:}\quad U(R)\,|\,k,\lambda\,\rangle & = & |\,Rk,\lambda\,\rangle\,\exp(-i\lambda w_{\mathrm{R}}(Rk\leftarrow k)),\nonumber \\
\mathrm{Boosts:}\quad U(\Lambda)\,|\,k,\lambda\,\rangle & = & |\,\Lambda k,\lambda\,\rangle\,\exp(-i\lambda w_{\mathrm{B}}(\Lambda k\leftarrow k)),\nonumber \\
\mathrm{Space\ inversion}:\quad U(\mathcal{P})\,|\,(\omega,\boldsymbol{k}),\lambda\,\rangle & = & |\,(\omega,-\boldsymbol{k}),-\lambda\,\rangle\,\eta\,e^{+i2\lambda\varphi_{k}},\nonumber \\
\mathrm{Time\ reversal}:\quad A(\mathcal{T})\,|\,(\omega,\boldsymbol{k}),\lambda\,\rangle & = & |\,(\omega,-\boldsymbol{k}),\lambda\,\rangle\,e^{-i2\lambda\varphi_{k}},\label{eq:14}
\end{eqnarray}
where $\hat{\boldsymbol{k}}=(\theta_{k},\varphi_{k})$ in spherical
polar coordinates and $\eta=-1$ is the intrinsic parity of the photon
\cite{Patrignani2016}. The Wigner rotations are calculated from
\begin{equation}
R_{z}(w_{\mathrm{R}}(Rk\leftarrow k))=R_{0}^{-1}[R\hat{\boldsymbol{k}}]\,R\,R_{0}[\hat{\boldsymbol{k}}]\label{eq:15}
\end{equation}
and
\begin{equation}
R_{z}(w_{\mathrm{B}}(\Lambda k\leftarrow k))\mathcal{L}(\boldsymbol{\alpha})=L^{-1}(\Lambda k,k_{0})\,\Lambda\,L(k,k_{0}).\label{eq:16}
\end{equation}
We find
\begin{equation}
\exp(-\frac{i}{2}w_{\mathrm{R}}(Rk\leftarrow k))=\frac{\mathcal{R}_{+\frac{1}{2},+\frac{1}{2}}\cos\frac{\theta_{k}}{2}+\mathcal{R}_{+\frac{1}{2},-\frac{1}{2}}\sin\frac{\theta_{k}}{2}\,e^{+i\varphi_{k}}}{\left|\mathcal{R}_{+\frac{1}{2},+\frac{1}{2}}\cos\frac{\theta_{k}}{2}+\mathcal{R}_{+\frac{1}{2},-\frac{1}{2}}\sin\frac{\theta_{k}}{2}\,e^{+i\varphi_{k}}\right|}\label{eq:16.1}
\end{equation}
and
\begin{equation}
\exp(-\frac{i}{2}w_{\mathrm{B}}(\Lambda k\leftarrow k))=\frac{(\cosh\frac{\zeta}{2}+\sinh\frac{\zeta}{2}\,\hat{\boldsymbol{\zeta}}\cdot\hat{\boldsymbol{z}})\cos\frac{\theta_{k}}{2}+\sinh\frac{\zeta}{2}\,\hat{\boldsymbol{\zeta}}\cdot(\hat{\boldsymbol{x}}-i\hat{\boldsymbol{y}})\,\sin\frac{\theta_{k}}{2}\,e^{+i\varphi_{k}}}{\left|(\cosh\frac{\zeta}{2}+\sinh\frac{\zeta}{2}\,\hat{\boldsymbol{\zeta}}\cdot\hat{\boldsymbol{z}})\cos\frac{\theta_{k}}{2}+\sinh\frac{\zeta}{2}\,\hat{\boldsymbol{\zeta}}\cdot(\hat{\boldsymbol{x}}-i\hat{\boldsymbol{y}})\,\sin\frac{\theta_{k}}{2}\,e^{+i\varphi_{k}}\right|},\label{eq:16.2}
\end{equation}
where
\begin{equation}
\mathcal{R}_{m_{1}m_{2}}=\langle\,\frac{1}{2},m_{1}\,|\,U(R)\,|\,\frac{1}{2},m_{2}\,\rangle\label{eq:16.3}
\end{equation}
and $\boldsymbol{\zeta}$ is the rapidity of the boost. These results
take special forms if the initial or final momentum direction is $\pm\hat{\boldsymbol{z}},$
but these points are a set of measure zero in the superposition integrals
that follow.

We construct normalized state vectors by
\begin{equation}
|\,\psi\,\rangle=\int\frac{d^{3}k}{\sqrt{\omega}}\sum_{\lambda=\pm1}|\,k,\lambda\,\rangle\Psi_{\lambda}(k).\label{eq:17}
\end{equation}
The factor of $1/\sqrt{\omega},$ as in the massive case \cite{Hoffmann2018d},
is to compensate for the invariant normalization. Then, just as we
did for massive particles, we find the transformation laws for the
quantities $\Psi_{\lambda}(k)$:
\begin{eqnarray}
\mathrm{Spacetime\ translations:}\quad\Psi_{\lambda}^{\prime}(k) & = & \Psi_{\lambda}(k)\,e^{+ik\cdot a},\nonumber \\
\mathrm{Rotations:}\quad\Psi_{\lambda}^{\prime}(k) & = & \Psi_{\lambda}(R^{-1}k)\,\exp(-i\lambda w_{\mathrm{R}}(k\leftarrow R^{-1}k)),\nonumber \\
\mathrm{Boosts:}\quad\Psi_{\lambda}^{\prime}(k) & = & \Psi_{\lambda}(\Lambda^{-1}k)\,\sqrt{\gamma_{0}(1-\boldsymbol{\beta}_{0}\cdot\hat{\boldsymbol{k}})}\,\exp(-i\lambda w_{\mathrm{B}}(k\leftarrow\Lambda^{-1}k)),\nonumber \\
\mathrm{Space\ inversion}:\quad\Psi_{\lambda}^{\prime}(\omega,\boldsymbol{k}) & = & \Psi_{-\lambda}(\omega,-\boldsymbol{k})\,\eta\,e^{+i2\lambda\varphi_{k}},\nonumber \\
\mathrm{Time\ reversal}:\quad\Psi^{\prime}(\omega,\boldsymbol{k}) & = & \Psi_{\lambda}^{*}(\omega,-\boldsymbol{k})\,e^{-i2\lambda\varphi_{k}}.\label{eq:18}
\end{eqnarray}

We note, as we noted in \cite{Hoffmann2018d}, that special relativity
does not require all quantities of physical interest to transform
as scalars, four-vectors or tensors. The requirements of special relativity
are met in these transformation properties, which allow the construction
of state vectors to describe a system in any frame.

The projector result,
\begin{equation}
\langle\,\psi\,|\,k,\lambda\,\rangle\langle\,k,\lambda\,|\,\psi\,\rangle=|\Psi_{\lambda}(k)|^{2},\label{eq:19}
\end{equation}
the normalization condition,
\begin{equation}
\int d^{3}k\sum_{\lambda=\pm1}|\Psi_{\lambda}(k)|^{2}=1\label{eq:20}
\end{equation}
and the average momentum formula
\begin{equation}
\langle\,\psi\,|\,P^{\mu}\,|\,\psi\,\rangle=\int d^{3}k\sum_{\lambda=\pm1}|\Psi_{\lambda}(k)|^{2}\,k^{\mu}\label{eq:21}
\end{equation}
confirm the interpretation of $\Psi_{\lambda}(k)$ as a momentum/helicity
probability amplitude.

\section{Localization of massless particles and the electromagnetic field}

For a hypothetical massless particle of zero helicity, the Newton-Wigner
construction \cite{Newton1949} gives localized state vectors
\begin{equation}
|\,\boldsymbol{x}\,\rangle=\int\frac{d^{3}k}{\sqrt{\omega}}\,|\,k\,\rangle\frac{e^{-i\boldsymbol{k}\cdot\boldsymbol{x}}}{(2\pi)^{\frac{3}{2}}}\label{eq:22}
\end{equation}
with
\begin{equation}
\langle\,\boldsymbol{x}_{1}\,|\,\boldsymbol{x}_{2}\,\rangle=\delta^{3}(\boldsymbol{x}_{1}-\boldsymbol{x}_{2}),\label{eq:23}
\end{equation}
just as in the massive case.

The impossibility of localized state vectors for the photon is not
because of its masslessness but because of its limited helicity spectrum.
If there were a $\lambda=0$ state for the photon, we could construct
three state vectors for particles localized at the origin
\begin{equation}
|\,\boldsymbol{0},\mu\,\rangle=\int\frac{d^{3}k}{\sqrt{\omega}}\sum_{\lambda=-1}^{1}|\,k,\lambda\,\rangle\mathcal{R}_{\lambda\mu}^{(1)-1}[\hat{\boldsymbol{k}}]\quad\mathrm{for}\ \mu=-1,0,+1,\label{eq:24}
\end{equation}
where $\mathcal{R}_{\lambda\mu}^{(1)}[\hat{k}]=\langle\,1,\lambda\,|\,U(R_{0}[\hat{\boldsymbol{k}}])\,|\,1,\mu\,\rangle$
are $J=1$ matrix elements of the standard helicity rotation, Eq.
(\ref{eq:11}). These three would satisfy Newton and Wigner's requirement
of rotating according to a finite-dimensional irreducible representation
of the rotation group:
\begin{equation}
U(R)\,|\,\boldsymbol{0},\mu\,\rangle=\sum_{\mu^{\prime}=-1}^{1}|\,\boldsymbol{0},\mu^{\prime}\,\rangle\mathcal{D}_{\mu^{\prime}\mu}^{(1)}(R).\label{eq:25}
\end{equation}
A spatial translation would then give state vectors orthogonal to
the ones localized at the origin. These two results have been obtained
previously by Wightman \cite{Wightman1962}.

\subsection{Electromagnetic field strength operators}

We clearly need at least a partial measure of localization for photons,
since they can be localized in a very small spatial volume with an
attosecond pulsed laser \cite{Li2017}. We seek this measure in expectation
values of the free electromagnetic field strength operators. From
quantum electrodynamics, for a system of an arbitrary number of free
photons, these are \cite{Itzykson1980}
\begin{equation}
F^{\mu\nu}(x)=\frac{1}{\sqrt{16\pi^{3}}}\int\frac{d^{3}k}{\omega}\sum_{\lambda=\pm1}(k^{\mu}\epsilon^{\nu}(k,\lambda)-k^{\nu}\epsilon^{\mu}(k,\lambda))\,a(k,\lambda)\,e^{-ik\cdot x}+(\dagger),\label{eq:25.1}
\end{equation}
where $(\dagger)$ is the Hermitian conjugate of the first term. (There
is a choice of overall sign here.) Note that we chose, for simplicity,
to form an Hermitian operator as the sum of a term and its Hermitian
conjugate. The more common convention is $i$ times the difference
of the term and its Hermitian conjugate. The difference is only a
phase shift in the spacetime dependence of the field. More importantly,
we note that in \cite{Itzykson1980}, the authors sum over \textit{four},
not two, photon polarizations. This is done to enforce locality of
the four-component, gauge-dependent potential $A^{\mu}(x)$ (see Eq.
(\ref{eq:56}) below), where $F^{\mu\nu}=\partial^{\mu}A^{\nu}-\partial^{\nu}A^{\mu}.$
The consequences are troubling, including a state vector with a negative
norm. Since we are not constructing a local quantum field theory in
this paper and we are dealing only with free photons, we feel justified
in using only the physical polarizations of the photon.

The annihilation and creation operators satisfy
\begin{equation}
[a(k_{1},\lambda_{1}),a^{\dagger}(k_{2},\lambda_{2})]=\delta_{\lambda_{1}\lambda_{2}}\,\omega_{1}\delta^{3}(\boldsymbol{k}_{1}-\boldsymbol{k}_{2}).\label{eq:25.2}
\end{equation}
The creation operators transform like
\begin{equation}
a^{\dagger}(k,\lambda)\sim|\,k,\lambda\,\rangle\langle\,0\,|,\label{eq:25.3}
\end{equation}
with the vacuum, $|\,0\,\rangle,$ invariant. The annihilation operators,
$a(k,\lambda),$ are the Hermitian conjugates of the creation operators.

Once the polarization vectors, $\epsilon^{\mu}(k,\lambda),$ are constructed,
which we do below, and their transformation properties known, it can
easily be seen that the field strengths transform locally under Lorentz
transformations, $L,$ as the elements of an antisymmetric tensor,
\begin{equation}
U^{\dagger}(L)\,F^{\mu\nu}(x)\,U(L)=L_{\phantom{\mu}\rho}^{\mu}L_{\phantom{\nu}\sigma}^{\nu}F^{\rho\sigma}(L^{-1}x),\label{eq:25.4}
\end{equation}
and that they satisfy the Maxwell equations
\begin{align}
\partial_{\mu}F^{\mu\nu}(x) & =0,\nonumber \\
\partial^{\rho}F^{\mu\nu}(x)+\partial^{\mu}F^{\nu\rho}(x)+\partial^{\nu}F^{\rho\mu}(x) & =0.\label{eq:25.5}
\end{align}

These fields are normalized in that
\begin{equation}
\int d^{3}x\,\Theta^{0\mu}(x)=\int\frac{d^{3}k}{\omega}\sum_{\lambda=\pm1}k^{\nu}a^{\dagger}(k,\lambda)a(k,\lambda)=P^{\mu},\label{eq:25.6}
\end{equation}
the total energy-momentum four-vector, where
\begin{equation}
\Theta^{\mu\nu}(x)=:\frac{1}{2}(F^{\mu\rho}(x)F_{\rho}^{\phantom{\rho}\nu}(x)+F^{\nu\rho}(x)F_{\rho}^{\phantom{\rho}\mu}(x))-\frac{1}{4}g^{\mu\nu}F^{\alpha\beta}(x)F_{\beta\alpha}(x):\label{eq:25.7}
\end{equation}
is the symmetric, traceless energy-momentum tensor as in classical
physics. Normal ordering (as indicated by the colons), with creation
operators to the left of annihilation operators, removes an unphysical
infinite term. (Proving the covariance of the integral, Eq. (\ref{eq:25.6}),
over all space involves integration over spacelike hypersurfaces \cite{Goldstein1980}
and will not be discussed here.) Note
\begin{equation}
\Theta^{0\mu}=:(\frac{1}{2}(\boldsymbol{E}^{2}+\boldsymbol{B}^{2}),\frac{1}{2}(\boldsymbol{E}\times\boldsymbol{B}-\boldsymbol{B}\times\boldsymbol{E}))^{\mu}:.\label{eq:25.8}
\end{equation}

\subsection{Construction of the polarization vectors}

We briefly review the construction of the polarization vectors $\epsilon^{\mu}(k,\lambda).$
We will see that there is some freedom, which we identify as gauge
freedom, in the definition. Our goal is to construct electromagnetic
field strengths that satisfy the Maxwell equations (Eqs. (\ref{eq:25.5}))
and transform locally as an antisymmetric tensor (Eq. (\ref{eq:25.4})).

If the polarization vectors satisfy the Lorentz condition,
\begin{equation}
k\cdot\epsilon(k,\lambda)=0,\label{eq:26}
\end{equation}
we see immediately that the Maxwell equations, Eqs. (\ref{eq:25.5}),
are satisfied. (More general schemes are possible, which we will not
consider.) We will see from Eq. (\ref{eq:34}) below that the Lorentz
condition is an invariant condition.

Any transformation of the polarization vector of the form
\begin{equation}
\epsilon^{\mu}(k,\lambda)\rightarrow\epsilon^{\mu}(k,\lambda)+f(k)\,k^{\mu}\label{eq:31.1}
\end{equation}
will give another polarization vector satisfying the Lorentz condition
(Eq. (\ref{eq:26})). The objects
\begin{equation}
T^{\mu\nu}(k,\lambda)=k^{\mu}\epsilon^{\nu}(k,\lambda)-k^{\nu}\epsilon^{\mu}(k,\lambda),\label{eq:35}
\end{equation}
appearing in $F^{\mu\nu}(x)$ (Eq. (\ref{eq:25.1})) are invariant
under such a transformation. So the transformation properties of the
electromagnetic field strengths and the satisfaction of the Maxwell
equations will be unchanged by such a transformation. Hence we have
gauge freedom in the definition of the polarization vectors. Any quantity
that depends on the polarization vectors but is not gauge invariant
must be considered unphysical, since there is not a unique definition
of such a quantity.

The construction begins with the definition of the polarization vectors
$\epsilon^{\mu}(k_{0},\lambda)$ for the reference states. The spatial
parts of the polarization vectors can be chosen, using a gauge transformation,
to be perpendicular to $\hat{\boldsymbol{z}}$. Then the zero components
must vanish to satisfy the Lorentz condition. For definiteness and
to make the rotation properties simple, we take the spatial parts
of the vectors to have components
\begin{equation}
\hat{\boldsymbol{i}}\cdot\boldsymbol{\epsilon}(k_{0},\lambda)=\sqrt{\frac{4\pi}{3}}\,Y_{1\lambda}(\hat{\boldsymbol{i}}).\label{eq:27}
\end{equation}
This gives
\begin{equation}
\epsilon^{\mu}(k_{0},+1)=-\frac{1}{\sqrt{2}}(0,1,i,0)^{\mu}\quad\mathrm{and}\quad\epsilon^{\mu}(k_{0},-1)=\frac{1}{\sqrt{2}}(0,1,-i,0)^{\mu}.\label{eq:28}
\end{equation}
These satisfy the complex orthonormality condition
\begin{equation}
\epsilon^{*}(k_{0},\lambda_{1})\cdot\epsilon(k_{0},\lambda_{2})=-\delta_{\lambda_{1}\lambda_{2}}.\label{eq:29}
\end{equation}

For the little group elements, we find
\begin{equation}
R_{z}(\gamma)_{\phantom{\mu}\nu}^{\mu}\epsilon^{\nu}(k_{0},\lambda)=\epsilon^{\mu}(k_{0},\lambda)\,e^{-i\lambda\gamma}\label{eq:30}
\end{equation}
and (using Eq. (\ref{eq:7}))
\begin{equation}
\mathcal{L}(\alpha)_{\phantom{\mu}\nu}^{\mu}\epsilon^{\nu}(k_{0},\lambda)=\epsilon^{\mu}(k_{0},\lambda)+\boldsymbol{\alpha}\cdot\boldsymbol{\epsilon}(k_{0},\lambda)\frac{k_{0}^{\mu}}{\kappa}.\label{eq:31}
\end{equation}
The rotation result is appropriate for the polarization vectors of
helicity eigenvectors. We see that it is not possible to construct
polarization vectors that are invariant under the IBR transformations.
Instead, the transformation involves the addition of a part proportional
to the four-momentum, which is a gauge transformation.

The polarization vectors for general momentum, $k,$ are then constructed
using the same Lorentz transformation as was used for the state vectors:
\begin{equation}
\epsilon^{\mu}(k,\lambda)=L(k,k_{0})_{\phantom{\mu}\nu}^{\mu}\epsilon^{\nu}(k_{0},\lambda)=R[\hat{\boldsymbol{k}}]_{\phantom{\mu}\nu}^{\mu}\epsilon^{\nu}(k_{0},\lambda),\label{eq:32}
\end{equation}
since the $z$ boost has no effect on the only transverse components.
Then we find the transformation properties
\begin{equation}
R_{\phantom{\mu}\nu}^{\mu}\epsilon^{\nu}(k,\lambda)=\epsilon^{\mu}(Rk,\lambda)\,\exp(-i\lambda w_{\mathrm{R}}(Rk\leftarrow k))\label{eq:33}
\end{equation}
and
\begin{equation}
\Lambda_{\phantom{\mu}\nu}^{\mu}\epsilon^{\nu}(k,\lambda)=\epsilon^{\mu}(\Lambda k,\lambda)\,\exp(-i\lambda w_{\mathrm{B}}(\Lambda k\leftarrow k))+\boldsymbol{\alpha}\cdot\boldsymbol{\epsilon}(k_{0},\lambda)\frac{(\Lambda k)^{\mu}}{\kappa},\label{eq:34}
\end{equation}
for an $\boldsymbol{\alpha}$ that can be calculated. We note that
this last equation gives
\begin{equation}
k^{\prime}\cdot\epsilon(k^{\prime},\lambda)=0,\label{eq:34.1}
\end{equation}
with $k^{\prime}=\Lambda k,$ so we see, as noted earlier, that the
Lorentz condition is an invariant statement.

In the derivation of the Lorentz transformation properties, Eq. (\ref{eq:25.4}),
the phase factor from
\[
U^{\dagger}(L)\,a(k,\lambda)\,U(L)=a(Lk,\lambda)\,\exp(+i\lambda w_{\mathrm{B}}(\Lambda k\leftarrow k))
\]
cancels with the phase factor from Eq. (\ref{eq:34}), and the local
covariance follows.

\subsection{Expectations of the electromagnetic field strengths in a coherent
state}

Because of the presence of the creation and annihilation operators
in Eq. (\ref{eq:25.1}), which change the photon number by $\pm1,$
respectively, the expectation value of the electromagnetic field strength
operators in a state of definite photon number vanishes. However we
can construct a coherent state \cite{Glauber1963a,Glauber1963b},
a superposition of all photon numbers with a mean number of unity.
The electromagnetic field strengths will have nonzero expectation
values in this coherent state, which expectation values we propose
as our measure of localization.

We first define
\begin{equation}
a(\psi)=\int\frac{d^{3}k}{\sqrt{\omega}}\sum_{\lambda=\pm1}\Psi_{\lambda}^{*}(k)\,a(k,\lambda)\label{eq:36}
\end{equation}
for a state vector $|\,\psi\,\rangle$ with probability amplitudes
$\Psi_{\lambda}(k)$ as in Eq. (\ref{eq:17}). The commutator with
its Hermitian conjugate is
\begin{equation}
[a(\psi),a^{\dagger}(\psi)]=\int\frac{d^{3}k_{1}}{\sqrt{\omega_{1}}}\sum_{\lambda_{1}=\pm1}\int\frac{d^{3}k_{2}}{\sqrt{\omega_{2}}}\Psi_{\lambda_{1}}^{*}(k_{1})\Psi_{\lambda_{2}}(k_{2})\delta_{\lambda_{1}\lambda_{2}}\omega_{1}\delta^{3}(\boldsymbol{k}_{1}-\boldsymbol{k}_{2})=\int d^{3}k\sum_{\lambda=\pm1}|\Psi_{\lambda}(k)|^{2}=1.\label{eq:37}
\end{equation}
Then a coherent state with arbitrary mean photon number (related to
$z$) is required to satisfy
\begin{equation}
a(\psi)\,|\,z,\psi\,\rangle=z\,|\,z,\psi\,\rangle.\label{eq:38}
\end{equation}
The solution is well known \cite{Glauber1963a,Glauber1963b}
\begin{equation}
|\,z,\psi\,\rangle=e^{-|z|^{2}/2}\sum_{n=0}^{\infty}\frac{(z\,a^{\dagger}(\psi))^{n}}{n!}\,|\,0\,\rangle.\label{eq:39}
\end{equation}

The photon number operator is
\begin{equation}
N=\int\frac{d^{3}k}{\omega}\sum_{\lambda=\pm1}a^{\dagger}(k,\lambda)a(k,\lambda).\label{eq:40}
\end{equation}
It is easily seen that
\begin{equation}
[N,a^{\dagger}(\psi)]=a^{\dagger}(\psi)\label{eq:41}
\end{equation}
and then
\begin{equation}
[N,(a^{\dagger}(\psi))^{n}]=n\,a^{\dagger}(\psi).\label{eq:42}
\end{equation}
So the expectation value of the number operator in our coherent state
is
\begin{equation}
\langle\,z,\psi\,|\,N\,|\,z,\psi\,\rangle=\langle\,z,\psi\,|\,z\,a^{\dagger}(\psi)\,|\,z,\psi\,\rangle=|z|^{2}.\label{eq:43}
\end{equation}
So we will take $z=1$ to make a coherent state that mimicks a single
photon state.

We need
\begin{align}
[a(k,\lambda),a^{\dagger}(\psi)] & =\sqrt{\omega}\,\Psi_{\lambda}(k),\nonumber \\{}
[a(k,\lambda),(a^{\dagger}(\psi))^{n}] & =n\sqrt{\omega}\,\Psi_{\lambda}(k),\label{eq:44}
\end{align}
which implies
\begin{align}
\langle\,1,\psi\,|\,a(k,\lambda)\,|\,1,\psi\,\rangle & =\sqrt{\omega}\,\Psi_{\lambda}(k)\,|z|^{2}=\sqrt{\omega}\,\Psi_{\lambda}(k),\nonumber \\
\langle\,1,\psi\,|\,a^{\dagger}(k,\lambda)\,|\,1,\psi\,\rangle & =\sqrt{\omega}\,\Psi_{\lambda}^{*}(k).\label{eq:45}
\end{align}

Then we find the expectation value of the electromagnetic field strengths,
Eq. (\ref{eq:25.1}), in this coherent state, with probability amplitude
chosen as
\begin{equation}
\Psi_{\lambda}(k)=\delta_{\lambda1}\,\frac{e^{-|\boldsymbol{k}-\boldsymbol{\kappa}|^{2}/4\sigma_{k}^{2}}}{(2\pi\sigma_{k}^{2})^{\frac{3}{4}}}\label{eq:46}
\end{equation}
for a state of only one helicity, with $\boldsymbol{\kappa}=\kappa\,\hat{\boldsymbol{z}}$
(not the same $\kappa$ as in Section II) and a sharply defined momentum:
$\sigma_{k}/\kappa\ll1.$ Slowly varying factors can be replaced by
their values at the wavefunction peak, giving
\begin{equation}
\langle\,1,\psi\,|\,F^{\mu\nu}(x)\,|\,1,\psi\,\rangle\cong\frac{i}{\sqrt{16\pi^{3}}}\frac{T^{\mu\nu}(\kappa,1)}{\sqrt{\kappa}}\int d^{3}k\,e^{-ik\cdot x}\,\frac{e^{-|\boldsymbol{k}-\boldsymbol{\kappa}|^{2}/4\sigma_{k}^{2}}}{(2\pi\sigma_{k}^{2})^{\frac{3}{4}}}+(*),\label{eq:47}
\end{equation}
where ($*$) is the complex conjugate of the first term and the fractional
error is $\mathcal{O}(\sigma_{k}/\kappa)$.

We expand
\begin{equation}
\omega(k)=\sqrt{(\boldsymbol{\kappa}+\boldsymbol{k}-\boldsymbol{\kappa})^{2}}=\kappa\sqrt{1+\frac{2\hat{\boldsymbol{z}}\cdot(\boldsymbol{k}-\boldsymbol{\kappa})}{\kappa}+\frac{(\boldsymbol{k}-\boldsymbol{\kappa})^{2}}{\kappa^{2}}}=\kappa+\hat{\boldsymbol{z}}\cdot(\boldsymbol{k}-\boldsymbol{\kappa})+\mathcal{O}(\frac{|\boldsymbol{k}-\boldsymbol{\kappa}|^{2}}{\kappa}).\label{eq:48}
\end{equation}
If we restrict our attention to times, $t$, with
\begin{equation}
|t|\ll\frac{\kappa}{\sigma_{k}}\sigma_{x},\label{eq:49}
\end{equation}
with $\sigma_{x}\sigma_{k}=1/2,$ we can ignore the correction terms
and spreading of the fields will be negligible.

Then the integral is \cite{Gradsteyn1980} (3.323.2)
\begin{equation}
\int d^{3}k\,e^{-ik\cdot x}\,\frac{e^{-|\boldsymbol{k}-\boldsymbol{\kappa}|^{2}/4\sigma_{k}^{2}}}{(2\pi\sigma_{k}^{2})^{\frac{3}{4}}}=\int d^{3}k\,e^{-i\kappa t}e^{i\boldsymbol{\kappa}\cdot\boldsymbol{x}}e^{-i\hat{\boldsymbol{z}}\cdot(\boldsymbol{k}-\boldsymbol{\kappa})t}e^{i(\boldsymbol{k}-\boldsymbol{\kappa})\cdot\boldsymbol{x}}\,\frac{e^{-|\boldsymbol{k}-\boldsymbol{\kappa}|^{2}/4\sigma_{k}^{2}}}{(2\pi\sigma_{k}^{2})^{\frac{3}{4}}}=(2\pi)^{\frac{3}{2}}\,e^{i\kappa(z-t)}\,\frac{e^{-|\boldsymbol{x}-\hat{\boldsymbol{z}}t|^{2}/4\sigma_{x}^{2}}}{(2\pi\sigma_{x}^{2})^{\frac{3}{4}}}.\label{eq:50}
\end{equation}

With
\begin{equation}
T^{\mu\nu}(\kappa,1)=\kappa\begin{pmatrix}0 &  & \epsilon^{j}\\
\\
-\epsilon^{i} &  & -\epsilon^{i}\delta_{j3}+\epsilon^{j}\delta_{i3}\\
\\
\end{pmatrix}\label{eq:51}
\end{equation}
and
\begin{equation}
F^{\mu\nu}=\begin{pmatrix}0 &  & -E_{j}\\
\\
+E_{i} &  & -\epsilon_{ijk}B_{k}\\
\\
\end{pmatrix},\label{eq:52}
\end{equation}
we find
\begin{align}
\bar{E}_{x}(x;\psi) & =\sqrt{\kappa}\,\frac{e^{-|\boldsymbol{x}-\hat{\boldsymbol{z}}t|^{2}/4\sigma_{x}^{2}}}{(2\pi\sigma_{x}^{2})^{\frac{3}{4}}}\,\cos(\kappa(z-t)),\nonumber \\
\bar{E}_{y}(x;\psi) & =-\sqrt{\kappa}\,\frac{e^{-|\boldsymbol{x}-\hat{\boldsymbol{z}}t|^{2}/4\sigma_{x}^{2}}}{(2\pi\sigma_{x}^{2})^{\frac{3}{4}}}\,\sin(\kappa(z-t)),\nonumber \\
\bar{B}_{x}(x;\psi) & =\sqrt{\kappa}\,\frac{e^{-|\boldsymbol{x}-\hat{\boldsymbol{z}}t|^{2}/4\sigma_{x}^{2}}}{(2\pi\sigma_{x}^{2})^{\frac{3}{4}}}\,\sin(\kappa(z-t)),\nonumber \\
\bar{B}_{y}(x;\psi) & =\sqrt{\kappa}\,\frac{e^{-|\boldsymbol{x}-\hat{\boldsymbol{z}}t|^{2}/4\sigma_{x}^{2}}}{(2\pi\sigma_{x}^{2})^{\frac{3}{4}}}\,\cos(\kappa(z-t)),\label{eq:53}
\end{align}
with the magnetic field always perpendicular to the electric field.

At a fixed position, the electric field vector viewed facing into
the oncoming wave rotates counter-clockwise with time, as appropriate
for a positive helicity (this is called a \textit{left} circularly
polarized wave) \cite{Jackson1975}. We see
\begin{equation}
\int d^{3}x\,\frac{1}{2}\{\bar{\boldsymbol{E}}(x;\psi)^{2}+\bar{\boldsymbol{B}}(x;\psi)^{2}\}=\kappa,\label{eq:54}
\end{equation}
equal to the average energy and
\begin{equation}
\int d^{3}x\,\bar{\boldsymbol{E}}(x;\psi)\times\bar{\boldsymbol{B}}(x;\psi)=\boldsymbol{\kappa}\label{eq:55}
\end{equation}
equal to the average momentum.

These field expectations give us the measure of localization for a
single photon that we were looking for. We know that electric and
magnetic fields Lorentz contract under boosts, so we can use a similar
argument to the one used for massive particles \cite{Hoffmann2018d}.
A photon can be localized in an arbitrarily small volume as viewed
from a boosted frame, without any physical change to the system.

For a blue photon ($\kappa=3.3\,\mathrm{eV}$) and $\sigma_{k}/\kappa=0.01,$
the above calculation gives the localization scale $\sigma_{x}=3.0\,\mu\mathrm{m}.$

We note that we can construct a four-component potential as
\begin{equation}
A^{\mu}(x)=\frac{i}{\sqrt{16\pi^{3}}}\int\frac{d^{3}k}{\omega}\sum_{\lambda=\pm1}\epsilon^{\mu}(k,\lambda)\,a(k,\lambda)\,e^{-ik\cdot x}+(\dagger),\label{eq:56}
\end{equation}
satisfying
\begin{equation}
F^{\mu\nu}(x)=\partial^{\mu}A^{\nu}(x)-\partial^{\nu}A^{\mu}(x)\label{eq:57}
\end{equation}
as in classical electrodynamics \cite{Jackson1975}. It is not gauge
invariant and only transforms like a four-vector up to a gauge transformation.

Sipe \cite{Sipe1995} constructed a three-vector function of spacetime
in one frame to be used as a ``position wavefunction'' for a single
photon. In our notation this is
\begin{equation}
\boldsymbol{\Psi}(x)=\int\frac{d^{3}k}{(2\pi)^{\frac{3}{2}}}e^{-ik\cdot x}\sum_{\lambda=\pm1}\boldsymbol{\epsilon}(k,\lambda)\,\sqrt{\omega}\,\Psi_{\lambda}(k).\label{eq:57.4}
\end{equation}
The intent was to make $\boldsymbol{\Psi}^{*}(x)\cdot\boldsymbol{\Psi}(x)$
something like an energy density. The integral over all space of this
proposed density is
\begin{equation}
\int d^{3}x\,\boldsymbol{\Psi}^{*}(x)\cdot\boldsymbol{\Psi}(x)=\int d^{3}k\sum_{\lambda=\pm1}|\Psi_{\lambda}(k)|^{2}\,\omega=\langle\,\psi\,|\,H\,|\,\psi\,\rangle.\label{eq:57.5}
\end{equation}
As written, the quantities $\boldsymbol{\Psi}(x)$ are not gauge invariant.
We could construct an alternative that is manifestly gauge invariant,
\begin{equation}
\tilde{\Psi}^{i}(x)=\int\frac{d^{3}k}{\omega(2\pi)^{\frac{3}{2}}}e^{-ik\cdot x}\sum_{\lambda=\pm1}(k^{0}\epsilon^{i}(k,\lambda)-k^{i}\epsilon^{0}(k,\lambda))\,\sqrt{\omega}\,\Psi_{\lambda}(k),\label{eq:57.6}
\end{equation}
in terms of general polarization vectors in any gauge satisfying the
Lorentz condition (\ref{eq:26}). This expression reduces to Eq. (\ref{eq:57.4})
for polarization vectors with $\epsilon^{0}(k,\lambda)=0.$

We identify these components as
\begin{equation}
\tilde{\Psi}^{i}(x)=\sqrt{2}\,F^{(+)0i}(x)=-\sqrt{2}\,E^{(+)i}(x),\label{eq:57.7}
\end{equation}
in terms of the positive frequency (energy) electric and magnetic
fields, obtained from
\begin{equation}
F^{(+)\mu\nu}(x)=\langle\,0\,|\,F^{\mu\nu}(x)\,|\,\psi\,\rangle=\frac{1}{\sqrt{16\pi^{3}}}\int\frac{d^{3}k}{\omega}\sum_{\lambda=\pm1}(k^{\mu}\epsilon^{\nu}(k,\lambda)-k^{\nu}\epsilon^{\mu}(k,\lambda))\,\sqrt{\omega}\,\Psi_{\lambda}(k)\,e^{-ik\cdot x}.\label{eq:57.8}
\end{equation}

The boost transformations of these quantities could now be easily
calculated and would involve positive energy magnetic fields, $\boldsymbol{B}^{(+)}(x)$,
not of the form (\ref{eq:57.6}). So just considering the positive
energy electric fields does not give a complete characterization of
the system. 

Bia\l ynicki-Birula \cite{Bialynicki-Birula1994} uses as ``wavefunctions''
for the photon
\begin{equation}
\boldsymbol{F}_{\pm}(x)=\frac{1}{\sqrt{2}}(\boldsymbol{E}^{(+)}(x)\pm i\boldsymbol{B}^{(+)}(x)),\label{eq:58}
\end{equation}
linear combinations of the positive energy electric and magnetic fields.
His measure of localization,
\begin{equation}
\rho(x)=F_{+}^{*}(x)\cdot F_{+}(x)+F_{-}^{*}(x)\cdot F_{-}(x),\label{eq:59}
\end{equation}
has the dimensions of an energy density and integrates over all space
to the expectation of energy, but is not equal to $\frac{1}{2}(\bar{\boldsymbol{E}}^{2}+\bar{\boldsymbol{B}}^{2})$.

Neither of these two ``wavefunctions'' is of the form
\begin{equation}
\psi_{\mu}(x)=\langle\,(t,\boldsymbol{x}),\mu\,|\,\psi\,\rangle,\label{eq:60}
\end{equation}
since we know that is not possible.

\section{Conclusions}

In coherent superpositions of the momentum/helicity eigenvectors for
a photon, we have identified momentum/helicity probability amplitudes
with a probability interpretation very similar to that for the momentum
wavefunctions of a massive particle. We have found the transformation
properties of these probability amplitudes under spacetime translations,
rotations, boosts, space inversion and time reversal.

With no position eigenvectors possible for the photon, we searched
for another measure of localization. The expectation values of the
electromagnetic field strength components were found in a wavepacket
coherent state with a mean photon number of unity. This led to expectations
partially localized in space. The classical integrals over all space
of $\frac{1}{2}(\bar{\boldsymbol{E}}^{2}+\bar{\boldsymbol{B}}^{2})$
and $\bar{\boldsymbol{E}}\times\bar{\boldsymbol{B}}$ gave the average
energy and the average momentum of the photon state, respectively.

\bibliographystyle{apsrev4-1}

\end{document}